\documentclass[10pt,onecolumn,oneside]{asme2ej}
\usepackage{pst-all}
\usepackage{amsmath,bm}
\usepackage{color}
\usepackage{cite}
\usepackage{psfrag}
\usepackage{graphicx}
\usepackage{subfig} 
\usepackage{hyperref}
\hypersetup{%
 bookmarks=false
pdftitle = {Dual~Time~Stepping~Algorithms~with~the~High~Order~Harmonic~Balance~Method~for~Contact~Interfaces~with~Fretting-Wear},
pdfauthor = {Salles~et~al.}
}

\usepackage{booktabs}

\DeclareGraphicsExtensions{.eps}
\graphicspath{{IMG/}}

\DeclareMathAlphabet\mbi{OML}{cmm}{b}{it} 
\newcommand{\vectg}[1]{\boldsymbol{\mathit{#1}}} 
\newcommand{\vect}[1]{\mbi{#1}}	
\newcommand{\mat}[1]{\mathbf{#1}}	

\definecolor{Vert}{rgb}{0.3,0.7,0.3}

\newcommand{\df}[2]{\frac{\partial #1}{\partial #2}}

\title{Dual Time Stepping Algorithms with the High Order Harmonic Balance Method for Contact Interfaces with Fretting-Wear}

\author{Lo\"{i}c Salles$^{a,b}$,
   \affiliation{
            ${}^{(a)}$  \'Ecole Centrale de Lyon, \\
            Laboratoire de Tribologie et Dynamique des Syst\`emes
            \\36 avenue Guy de Collongue, 69134 Ecully Cedex, France\\
        ${}^{(b)}$ Moscow State Technical University named after Bauman, \\
        Moscow, Russia \\
      \\
      \texttt{\small Loic.Salles@ec-lyon.fr}\hspace{1cm}
   }
}

\author{ Laurent Blanc,Fabrice Thouverez
   \affiliation{
           \'Ecole Centrale de Lyon, \\
           Laboratoire de Tribologie et Dynamique des Syst\`emes
            \\36 avenue Guy de Collongue, 69134 Ecully Cedex, France\\
   }
}

\author{Alexander M. Gouskov
   \affiliation{
       Moscow State Technical University named after Bauman, \\
       2nd bauman Ul.5,105005 Moscow, Russia 
   }
}

\author{Pierrick Jean
   \affiliation{
             Snecma -- Safran group, \\
            77550 Moissy-Cramayel, France\\
   }
}

\usepackage{setspace}

\begin{document}

\maketitle    

\begin{abstract}
{\it Contact interfaces with dry friction are frequently used in turbomachinery. 
Dry friction damping produced by the sliding surfaces of these interfaces reduces the amplitude of bladed-disk vibration. The relative displacements at these interfaces lead to fretting-wear which  reduces the average life expectancy of the structure. 
Frequency response functions are calculated numerically by using the multi-Harmonic Balance Method (mHBM). The Dynamic Lagrangian Frequency-Time method is used to calculate contact forces in the frequency domain. A new strategy for solving non-linear systems based on dual time stepping is applied. 
This method is faster than using Newton solvers. 
It was used successfully for solving Nonlinear CFD equations in the frequency domain. This new approach allows identifying the steady state of worn systems by integrating wear rate equations a on dual time scale. The dual time equations are integrated by an implicit scheme. 
Of the different orders tested, the first order scheme provided the best results.}
\end{abstract}

\newpage
\begin{nomenclature}
    \entry{$\tilde{\text{\textbullet}}$}{Multiharmonic vectors.}
\entry{$\bar{\text{\textbullet}}$}{Retained time step value}
\entry{${\_}^T, {\_}^N$}{Tangential and normal direction.}
\entry{$p_N,p_n$}{Normal pressures.}
\entry{$p_T,p_T$}{Tangential shear.}
\entry{$w,\vect{W}$}{Wear depth, nodal wear depth vector.}
\entry{${\_}_r$}{Relative value.}
\entry{$\vect{U}$}{Time-domain nodal displacement vector.}
\entry{$\mat{M}$, $\mat{C}$, $\mat{K}$}{Mass, viscous damping and stiffness matrices.}
\entry{$\mat{Z}_r$}{Reduced dynamic stiffness on relative displacements.}
\entry{$F_c$}{Nodal force of contact.}
\entry{$\bar{\vectg{\lambda}}$, $\tilde{\vectg{\lambda}}$} {Lagrangian multiplier in time and frequency domains.}
\end{nomenclature}

\section*{INTRODUCTION}
Sources of non-linearity in turbomachinery are multiple.
Contact with friction is one of the most important causes of non-linearity in turbomachinery, especially in bladed-disks.
Contact with friction must be taken into account to ensure efficient that the vibration of this type of structure can be predicted well.
Friction dampers can be added to decrease response to vibration, although this requires numerical tools to evaluate the added damping rate.
In bladed disks, contact with friction occurs at the interfaces between the blades and the disk to which they are attached.
Fluid is another source of non-linearity.
It is possible to simulate the periodic behaviour of coupled fluid-structures by using the frequency method.

HBM (Harmonic Balance Method) is the most widely used frequency method. 
It is based on the expansion of variables in Fourier series and the Galerkin procedure to obtain non-linear algebraic systems.
For systems with contact and friction, an alternating frequency time (AFT) procedure is performed to calculate non-linear forces in the time domain and then transform them into the frequency domain.

Other approaches are possible, in this paper we study two other methods: the trigonometric collocation method (TCM) and the high dimension harmonic balance method (HDHB).
In TCM a non-linear algebraic system is solved in the time domain.
In the HDHB method the unknowns are values of displacements with an equal time step and the non-linear algebraic system is solved once again in the time domain.
Some authors have called this method the time spectral method (TSM)~\cite{gopinath2005} and it was used in CFD and proved highly efficient. 

In the first section we present the theory underlying these methods and explain the advantages of each one.

Different strategies are possible for solving non-linear algebraic systems the most common of which is Newton-Raphson solver~\cite{ferri1986}.
For very large systems a Jacobian matrix can be ill-conditioned, with problems of convergence in the linear direction search step.
For CFD applications, several authors~\cite{jameson2001,gopinath2005,sicot2008} propose transforming the non-linear algebraic system into a first order differential system and integrating it to obtain the steady state and thus the solution of the non-linear system.
The second section sets out the theory of this method and different schemes are tested by using numerical examples.

In previous papers~\cite{salles2009,salles2010a,salles2010b} we have shown that steady state in fretting-wear can occur during vibration.
To identify steady state we proposed an integration scheme involving the calculation of transient kinetics.
A method that allows finding the steady state directly would be very useful.
In this paper we propose using a pseudo-time method and integrating wear kinetics in pseudo-time.
It is shown that the worn profiles obtained by using this method are the same as worn profiles obtained previously by numerical schemes coupled with the DLFT method~\cite{nacivet2003,salles2010b}.

\section{HB THEORY}
A general elastic structure with $Q$ degrees of freedom is considered. 
The linear structural model can be derived by using the finite element method or any other method, such as component mode synthesis, but all the degrees of freedom where contact with friction takes place must be retained as physical coordinates.
A contact element is defined as a set of two nodes between which contact occurs. 
For such a structure, the equations of motion can be written as:
\begin{equation}
    \mat{M} \ddot{\vect{U}} + \mat{C} \dot{\vect{U}} + \mat{K} \left( \vect{U}, \Omega \right) \vect{U} + \vect{F}_{c}(\vect{U} , \dot{\vect{U}} ) =   \vect{F}_{ex}(t) 
    \label{eq:motion}
\end{equation}

where the vectors  $\ddot{\vect{U}}$ , $\dot{\vect{U}}$ and  $\vect{U}$ are, respectively, the displacement, velocity, and acceleration of the structure, $ \vect{U}$ is the vector of displacements, $ \mat{M}, \mat{C}, \mat{K}$ are the mass matrix, viscous damping matrix and stiffness matrix, respectively.
$ \vect{F}_{ex}(t)$ is the vector of the external forces (periodic excitation at frequency $\omega$), and $\vect{F}_{c}$ represents the non-linear contact forces due to friction.

For the bladed-disk system, $\mat{K}$ contains centrifugal and geometric stiffnesses that are linked to the speed of rotation.

By assuming a steady state periodic response,  $\vect{U}$ can be written as a Fourier series: 
\begin{equation}
	\vect{U}(\tau)=\vect{\tilde{U}}_0 + \sum_{n=1}^{Nh} \vect{\tilde{U}}_{n,c} \cos ( n \tau ) + \vect{\tilde{U}}_{n,s} \sin ( n \tau)
	\label{eq:fourier}
\end{equation}
where $Nh$ is the number of temporal harmonics retained and $\tau=\omega t$ is the normalized time of the vibration period.

Eq.~\eqref{eq:fourier} is introduced in \eqref{eq:motion}. 
The following paragraph presents three different approaches for expressing equivalent non-linear systems to obtain steady state.

\subsection{Harmonic balance method with the AFT procedure}

\begin{figure}
   \begin{center}
        \psfrag{U}{$\tilde{U}$}
        \psfrag{FNL}{$\tilde{F}_{NL}$}
        \psfrag{u}{$u(t_k)$ et $\dot{u}(t_k)$}
        \psfrag{f(t,u,du)}{$f_{nl}(t_k,u(t_k),\dot{u}(t_k))$}
        \psfrag{iDFT}{iDFT}
        \psfrag{DFT}{DFT}
        \psfrag{domaine frequentiel}{ \quad frequency domain}        
        \psfrag{domaine temporel}{\quad time domain}
\includegraphics[width=0.8\columnwidth]{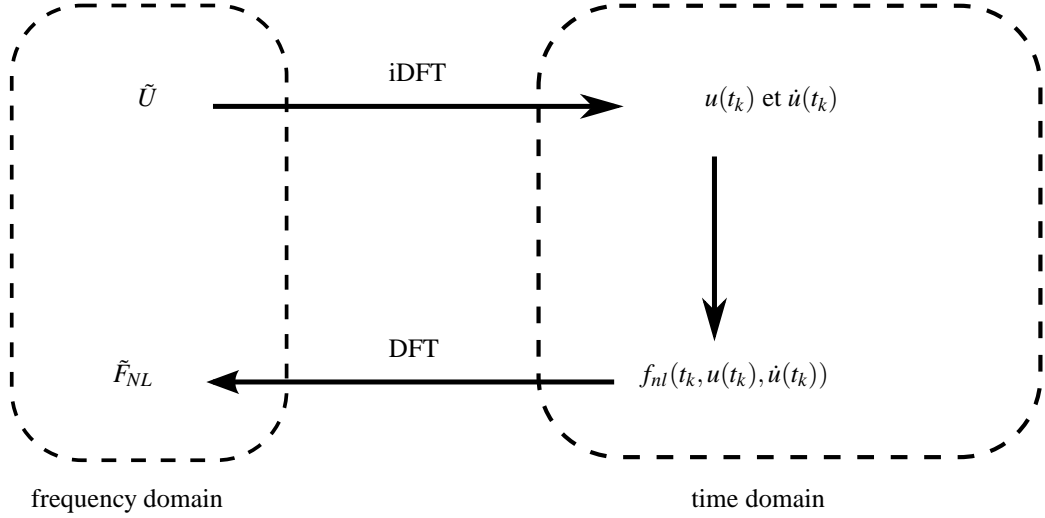}
    \end{center}
    \caption{AFT procedure}
    \label{fig:aft}
\end{figure}

In this method the unknowns are Fourier coefficients: $\tilde{\vect{U}} = \left( \vect{\tilde{U}}_0,\cdots, \vect{\tilde{U}}_{Nh,c} \vect{\tilde{U}}_{Nh,s}  \right)^T$.
Then the equation of motion~\eqref{eq:motion} can be written in the frequency domain by using Galerkin's procedure.
Thus the system is solved in the frequency domain with HBM.
The non-linear frequency algebraic system is condensed for non-linear dofs and then for relative displacements~\cite{nacivet2003} $\vect{\tilde{U}_r}$:
\begin{equation}
	\mat{Z_r} \vect{\tilde{U}_r}+\vectg{\tilde{F}}_c=\vect{\tilde{F}_r}
	\label{eq:hbm}
\end{equation}
$\mat{Z_r}$ is condensed dynamic stiffness and derived from dynamic stiffness $\mat{Z}$, which is a block diagonal matrix:
\begin{equation}                        
    \mat{Z}= \left[
    \begin{array}{cccc}
        \mat{K} & \cdots  & \mat{0}                        & \mat{0} \\
        \vdots  & \ddots  &  \mat{0}                        & \mat{0}\\
        \mat{0} & \mat{0} & \mat{K} - (N_h \omega)^2 \mat{M} & N_h \omega \mat{C} \\
        \mat{0} & \mat{0} & - N_h \omega \mat{C}              &  \mat{K} - (N_h \omega)^2 \mat{M} 
    \end{array}
    \right]                
\end{equation}
$\vect{\tilde{F}_r}$ is condensed multiharmonic vector of excitation forces.
Contact forces $\vectg{\tilde{F}}_c$  cannot be explicitly handled in the frequency domain.
Therefore, non-linear contact and friction forces are calculated by using a time-marching procedure in the time domain.
The AFT procedure is applied and the variables are transformed from the frequency domain to the time domain and contact forces are calculated.
Then they are transformed in the frequency domain as illustrated in Fig.~\ref{fig:aft}.
In this work the DLFT (Dynamic Lagrangian Frequency Time) method~\cite{nacivet2003, salles2009} is applied to calculate the Fourier coefficients of contact forces.
This method is based on a prediction-correction procedure with prediction in the frequency domain and correction to ensure contact with friction conditions (separation, stick, slip) in the time domain.
Other methods can be used as penalty methods~\cite{petrov2004,laxalde2007}, augmented lagrangian~\cite{laxalde2009} or methods based on more complicated contact laws (Dahl, Bouc-Wen\dots).

The advantages of HBM with the AFT procedure is that matrix $\mat{Z_r}$ is a block diagonal matrix and so, during the condensation step, it is possible to invert the matrices of each harmonic at the same time in parallel, so reducing the size of the inverse matrices.
On the other hand, during the AFT procedure, transformation in the time domain and return into the frequency domain must be performed, which is time consuming.
The number of time marching steps has no influence on the size of the non-linear system and  no aliasing problem occurs if the spectrum of contact forces in the time domain contains numerous harmonics ~\cite{labryer2009a,labryer2009b}. 

\subsection{Trigonometric collocation method}

Fourier coefficients are used again in this approach.
Instead of solving the system in the frequency domain, it is solved in the time domain.
The main idea is to take $2 Nh+1$ equally distributed time steps, which gives $(2 Nh+1) Q$ equations for $(2 Nh+1) Q$ unknowns.
It is possible to take more time steps; in this case the least squares method is applied.
The DLFT method is applied once again to calculate the contact forces.
When correcting the time domain, the number of time steps can be choosen as greater than $2 Nh+1$, then $2 Nh+1$ time steps will be chosen to express the following non-linear equation:
\begin{equation}
    f(\tilde{\vect{U}}) = \left\lbrace
    \begin{array}{llll}
           \mat{M} \ddot{\vect{U}}(\tau_1) + \mat{C} \dot{\vect{U}}(\tau_1) + \mat{K} \vect{U}(\tau_1) + \vect{F}_{c}(\tau_1) - \vect{F}_{ex}(\tau_1)\\
           \cdots\\
           \mat{M} \ddot{\vect{U}}(\tau_k) + \mat{C} \dot{\vect{U}}(\tau_k) + \mat{K} \vect{U}(\tau_k) + \vect{F}_{c}(\tau_k) - \vect{F}_{ex}(\tau_k)\\
           \cdots \\
           \begin{split}
           \mat{M} \ddot{\vect{U}}(\tau_{2Nh+1}) + \mat{C} \dot{\vect{U}}&(\tau_{2Nh+1}) + \mat{K} \vect{U}(\tau_{2Nh+1}) +\\ 
           &\vect{F}_{c}(\tau_{2Nh+1}) - \vect{F}_{ex}(\tau_{2Nh+1})
           \end{split}
    \end{array}
 \right.
    \label{eq:col}
\end{equation}
where $\tau_k$ are retained time steps ($k=[1:2*Nh+1]$ and $\vect{U} ( \tau_k ) = \vect{\bar{U}}_k =\displaystyle \vect{\tilde{U}}_0 + \sum_{n=1}^{Nh} \vect{\tilde{U}}_{n,c} \cos ( n \tau_k ) + \vect{\tilde{U}}_{n,s} \sin ( n \tau_k)$.

It is preferable to write these equations in another way by using operator formalism.
Eq.~\eqref{eq:fourier} can be rewritten as follows:
\begin{equation}
    \bar{\vect{U}} = \mat{T}^{-1} \vect{\tilde{U}}
    \label{eq:fourier2}
\end{equation}
where $ \mat{T}$ is a matrix of the discrete Fourier transform whose size is equal to $(2 Nh+1) Q$ by $(2 Nh+1) Q$. 
$\bar{\vect{U}}$ is the vector of all the displacements at all the selected time steps.
Expression of $\mat{T}$ is:
\begin{equation}
    \mat{T} = \frac{2}{2 Nh+1} \left[
    \begin{array}{cccc}
        \frac{1}{2} &  \frac{1}{2} & \dots & \frac{1}{2} \\
        \cos \tau_0 & \cos \tau_1 & \dots & \cos \tau_{2*Nh+1} \\
        \sin \tau_0 & \sin \tau_1 & \dots & \sin \tau_{2*Nh+1} \\
        \dots\\
        \cos Nh\tau_0 & \cos Nh\tau_1 & \dots & \cos Nh\tau_{2*Nh+1} \\
        \sin Nh\tau_0 & \sin Nh\tau_1 & \dots & \sin Nh\tau_{2*Nh+1}
    \end{array}
    \right] \otimes \mat{I}_N
    \label{eq:T}
\end{equation}

By using the dynamic stiffness matrix of the HBM problem, Eq.~\eqref{eq:col} is equivalent to:
\begin{equation}
    f(\tilde{\vect{U}}) = \mat{T}^{-1} \mat{Z}_r  \vect{\tilde{U}} + \bar{\vect{F}}_c - \vect{\bar{F}}_r
    \label{eq:col_Z}
\end{equation}

As the non-linear system is solved in the time domain, the contact forces are "exact" values.
In the classical HB method, the Fourier transform smooths the contact forces.

\subsection{High dimension harmonic balance method}

The last approach dealt with is the high dimension harmonic balance method proposed by Hall~\cite{hall2002} for CFD applications.
Liu~\cite{liu2006,liu2007} and LaBryer~\cite{labryer2009b,labryer2010} have also studied this method.
Some authors~\cite{sicot2008,gopinath2005} call this method the time spectral method.
In this paper we call this approach the high dimension harmonic balance method since it appears suitable for high dimensional dynamical systems.
The key aspect of the HDHB method is that instead of having Fourier coefficients as unknowns, as in the conventional HB approach, the unknowns are cast in the time domain and stored at $2Nh + 1$ equally spaced sub-time levels over a period of one cycle of motion.

As in the case of the trigonometric collocation method, the Fourier coefficients and time variables retained are linked by:
\begin{equation}
    \vect{\tilde{U}} = \mat{T} \bar{\vect{U}} \quad \text{and} \quad  \bar{\vect{U}} = \mat{T}^{-1} \vect{\tilde{U}}
    \label{eq:hdhb_T}
\end{equation}
By introducing expressions of Eq.~\eqref{eq:hdhb_T} in  Eq.~\eqref{eq:hbm} the following non-linear system is obtained:
\begin{equation}
    \mat{H}_r \bar{\vect{U}} + \bar{\vect{F}}_c = \bar{\vect{F}}_r
    \label{eq:hdhb}
\end{equation}
where $\displaystyle \mat{H}_r =  \mat{T} \mat{Z}_r  \mat{T}^{-1}$, this matrix is full rather than block diagonal.
The link between time events is ensured by $\mat{H}_r$.
The HDHB method does not require an AFT procedure to calculate the contact forces.
Any method used in a static contact problem can be used to solve Eq.~\eqref{eq:hdhb}.
In this paper we use a dynamic lagrangian method~\cite{nacivet2003}.

One of the problems with the HDBD method is that the number of time steps can only be $2Nh+1$.
Aliasing can occur if non-linearity introduces a higher harmonic order.
LaBryer described this phenomenom in ~\cite{labryer2010} and proposed filters to decrease aliasing.
The numerical examples given here show that aliasing influences the convergence of the method, especially for the low harmonic orders retained.  
\section{CALCULATION OF CONTACT FORCES}
Explanations of the DLFT method are given in following paragraph.

For HBM and TCM, the Lagrange multiplier $\vectg{\lambda}$ (corresponding to contact forces) is formulated as a penalization of the equations of motion in the tangential and normal directions in the frequency domain:
\begin{subequations}
\begin{eqnarray}
	\vectg{\tilde{\lambda}^T}&=&\vect{\tilde{F}_r}^T-\mat{Z_r} \vect{\tilde{U}_r}+\epsilon_T \left( \vect{\tilde{U}_r}^T-\vect{\tilde{X}_r}^T \right),\\
	\vectg{\tilde{\lambda}^N}&=&\vect{\tilde{F}_r}^N-\mat{Z_r} \vect{\tilde{U}_r}+\epsilon_N \left( \vect{\tilde{U}_r}^N-\vect{W}-\vect{\tilde{X}_r}^N \right),
\end{eqnarray}
\label{eq:lagrangian}
\end{subequations}
$\epsilon_T$ and $\epsilon_N$ are penalty coefficients, $\vect{\tilde{X}_r}$ is a new vector of relative displacements, which is computed in the time domain
and it corresponds to the relative displacements satisfying the contact and friction laws in the time domain.
$\vect{W}$ is the vector of the wear depths if fretting-wear occurs at the contact interfaces.
For the HDHB method, the equivalent expressions at Eq.~\eqref{eq:lagrangian} are written as:
\begin{subequations}
\begin{eqnarray}
	\vectg{\bar{\lambda}^T}&=&\vect{\bar{F}_r}^T-\mat{H_r} \vect{\bar{U}_r}+\epsilon_T \left( \vect{\bar{U}_r}^T-\vect{\bar{X}_r}^T \right),\\
	\vectg{\bar{\lambda}^N}&=&\vect{\bar{F}_r}^N-\mat{H_r} \vect{\bar{U}_r}+\epsilon_N \left( \vect{\bar{U}_r}^N-\vect{W}-\vect{\bar{X}_r}^N \right),
\end{eqnarray}
\label{eq:hdhb_lagrangian}
\end{subequations}
Eq.~\eqref{eq:lagrangian} and Eq.~\eqref{eq:hdhb_lagrangian} will be named Global step of DLFT method.

In time domain, contact forces are predicted by considering contact with a stick step. 
The period is split into $nit$ time steps.
In the time domain each vector of the global step has a counterpart. $\vectg{\tilde{\lambda}}$,$\vectg{\tilde{\lambda}_u}$ and $\vectg{\tilde{\lambda}_x}$ have respectively $\left\{ \vectg{\lambda^k} \right\}_{k=1..{nit}}$, $\left\{ \vectg{\lambda_u^k} \right\}_{k=1..{nit}}$ and $\left\{\vectg{\lambda_x^k} \right\}_{k=1..{nit}}$ as local equivalents. 
For the HDHB method, the global and local variables are the same, since the system is solved in the time domain with time variables as unknowns ($\left\{ \vectg{\lambda^k} \right\}_{k=1..{nit}} = \bar{\lambda}$, $\left\{ \vectg{\lambda_u^k} \right\}_{k=1..{nit}} = \bar{\lambda}_u$ and $\left\{\vectg{\lambda_x^k} \right\}_{k=1..{nit}} = \bar{\lambda}_x$ . 
For TCM and HBM $nit$ can be arbitrarily chosen; however for the HDHB method $nit=2*Nh+1$.

A prediction/correction is used to compute the contact forces. 
At each time increment it is assumed that the contact node is in stick situation, thus the node does not move and $\vectg{\lambda_x^{k,T}}=\vectg{\lambda_x^{k-1,T}}$ and $\vectg{\lambda_x^{k,N}}=0$. 
The predicted contact forces are:
\begin{equation}
\vectg{\lambda_{pre}^{k,T}} = \vectg{\lambda_u^{k,T}}-\vectg{\lambda_x^{k-1,T}}, \quad \lambda_{pre}^{k,N} = \lambda_u^{k,N}.
\label{eq:predi}
\end{equation} 
The corrected contact forces will be:
\begin{equation}
\vectg{\lambda^k} = \vectg{\lambda_u^k}-\vectg{\lambda_x^k}
\label{eq:corre}
\end{equation}
$\vectg{\lambda_x^k}$ will be calculated to satisfy the contact and friction laws.

\begin{enumerate}
	\item Separation: $\lambda_{pre}^{k,N} \ge 0$, contact is lost and the forces should be zero
\begin{equation}
\vectg{\lambda_x^k}=\vectg{\lambda_u^k},
\label{eq:Correction_sep}
\end{equation}
	\item Stick: $\lambda_{pre}^{k,N} < 0$ and $ \left\| \vectg{\lambda_{pre}^{k,T}} \right\| < \mu \left| \lambda_{pre}^{k,N} \right|$\\
In this case, the prediction verifies the contact conditions:
\begin{equation}
	\lambda_x^{k,N}=0, \quad \vectg{\lambda_x^{k,T}}=\vectg{\lambda_x^{k-1,T}},
\end{equation}
	\item Slip: $\lambda_{pre}^{k,N} < 0$ and $ \left\| \vectg{\lambda_{pre}^{k,T}} \right\| \ge \mu \left| \lambda_{pre}^{k,N} \right|$\\  
Again, there is no normal relative displacement. The correction is performed by assuming that the tangential contact force has the same direction as the predicted tangential force. The definition of relative velocity and compliance with Coulomb's law lead to:
\begin{equation}
\lambda_x^{k,N}=0, \quad 
\vectg{\lambda_x^{k,T}} = \vectg{\lambda_x^{k-1,T}} + \vectg{\lambda_{pre}^{k,T}} \left( 1-\mu\frac{|\lambda_{pre}^{k,N}|}{\|\vectg{\lambda_{pre}^{k,T}}\|} \right).
\label{eq:Correction_slip}
\end{equation}
\end{enumerate}

The final step consists of back-transforming the time domain updated lagrangian in the frequency domain by using a DFT algorithm for the Harmonic Balance Method, by taking the contact forces at the time steps chosen for the Trigonometric Balance Method, and by taking the vector of contact forces through time for the HDHB method.

\section{NON-LINEAR SOLVERS}
The solution of the HDHB, HBM and TCM system can be obtained numerically by using an iterative root finding scheme such as the Newton-Raphson method or Powell's hybrid method ~\cite{powell1970}.
In~\cite{salles2009} we proposed an analytical expression of a contact force gradient calculated in the time domain that allows faster calculation of the Jacobian matrix than with the finite difference method.

An original approach has been proposed for the CFD problem.
Obtaining the steady state of the RANS problem defined by a non-linear algebraic system can be transformed into a non-linear first order differential equation.
The method is used to solve a URANS problem expressed with HBM~\cite{mcmullen2006} or TSM method~\cite{sicot2008}.
The application of this strategy in Eq.~\eqref{eq:hdhb} in the case of the HDHB method gives the following system to be integrated:
\begin{equation}
    \df{\bar{\vect{U}}}{t^*} + \mat{H}_r \bar{\vect{U}} + \bar{\vect{F}}_c = \bar{\vect{F}}_r   
    \label{eq:pseudo}
\end{equation}
where $t^*$ is a pseudo-time that allows solving Eq.~\eqref{eq:hdhb}.
At the convergence of $ \df{\bar{\vect{U}}}{t^*} = 0$, which gives the solution of  Eq.~\eqref{eq:hdhb}.

This strategy converges faster than the classical Newton-Raphson method for the system with a large number of degrees of freedom.
Moreover, for large systems the Jacobian matrix can be ill-conditioned which leads to difficulties for solving the linear system during the direction search step with the Newton-Raphson method.

With the application of CFD, a different pseudo-time step can be chosen for each fluid cell.
For time integration, the time step is usually driven by the smallest cell. 
The transfer of information between all the cells is very slow,
but implementing different time steps along the mesh ensures optimal transfer of the information.
Regarding structural applications, this procedure can be adapted to choose different time-steps.
However, this strategy was not implemented here but will be in a future study.

The pseudo-time step can be fixed by using CFL (Courant–Friedrichs–Levy) condition:
\begin{equation}
    \Delta t^* =C  \frac{\Delta l}{c_0}
    \label{eq:CFL}
\end{equation}
where $C$ is a real coefficient, $\Delta l$ smallest length of cell and $c_0$ wave speed in the cell. 

Information is given in the numerical examples about different schemes in MATLAB, corresponding to more common numerical explicit and implicit schemes.
\section{FRETTING-WEAR WITH PSEUDO-TIME}
A study of fretting-wear in dovetail attachments during blade-disk vibration was described in a previous paper~\cite{salles2009,salles2010a} and it was shown that steady state can occur.
Therefore it would be useful to have a method capable of calculating the steady state caused by fretting-wear under dynamical loading directly.

Previously, the non-linear algeabric system was transformed into a differential equation introducing pseudo-time. This poses a problem of evolution, and fretting-wear is also an evolution problem defined by the differential equation:
\begin{equation}
    \dot{w} = k_w p_N \| \dot{u}_T \|
    \label{eq:archard}
\end{equation}
In~\cite{salles2010a} it was shown that fretting-wear during vibration can be considered as a problem with two time scales: fast time scales for vibration and long time scales for wear-kinetics.
Wear is assumed to be constant during one period.
In line with this logic, the pseudo-time scale will be considered as a slow time scale.
The wear kinetics equation is written as follows:
\begin{equation}
    \df{w}{t^*} =  k_w \int_0^{\frac{2 \pi}{\omega}} p_N(\tau) \| \dot{u}_T(\tau) \| d\tau
    \label{eq:wear_pseudo}
\end{equation}
The following augmented system must be integrated:
\begin{subequations}
\begin{eqnarray}
    \df{\bar{\vect{U}}}{t^*} &=& \mat{H}_r \bar{\vect{U}} + \bar{\vect{F}}_c - \bar{\vect{F}}_r\\
    \df{\vect{W}}{t^*} &=&  k_w \sum_{n=1}^{nit} \bar{\vect{P}}_N(\tau_n, \vect{W}) \| \bar{\dot{\vect{u}}}_T(\tau_n,\vect{W}) \| \Delta \tau 
    \label{eq:fretting_pseudo}
\end{eqnarray}
\end{subequations}
where $\Delta \tau = \tau_n - \tau_{n-1}$ is time step between two instancts.
In steady state, the solution of the dynamical problem is obtained and the wear rate is null. This is caracteristic of the steady state of the fretting-wear problem under dynamical loading.

This method will be illustrated by the numerical example studied in~\cite{salles2010b}.

\section{NUMERICAL EXAMPLES}

\begin{figure}
\begin{center}
\includegraphics[width=0.27\columnwidth]{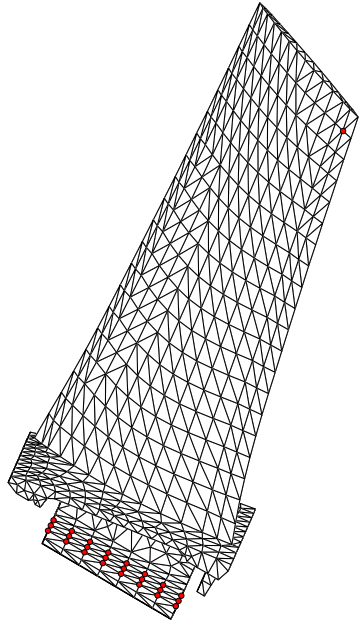}
\label{fig:aube}
\caption{Calculated bladed sector: finite element geometry}
\end{center}
\end{figure}

\begin{figure}
\begin{center}
\includegraphics[width=0.66\columnwidth]{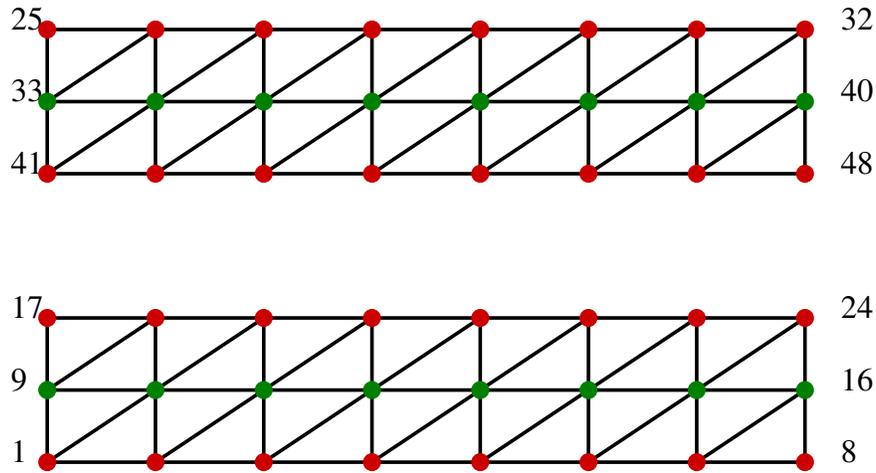}
\caption{Contact nodes on the intrados (top) and on the extrados (bottom)}      
\label{fig:noeuds}
\end{center}
\end{figure}

The method proposed will be tested and compared on a bladed-disk system (Fig.~\ref{fig:aube}).
Contact with friction occurs in the dovetail attachment.
The disk is made of 47 sectors.
In this paper only one sector with a fixed boundary is modeled.
In Fig.~\ref{fig:noeuds} the upper grid shows the numbering for the contact area on the same side as the intrados of the blade while the lower grid shows the numbering for the extrados side of the blade. 
The left side of the grid is the leading edge of the blade while the right side is its trailing edge.  
The friction coefficient is $\mu = 0.5$, corresponding to a contact between parts made of titanium without coating.
The first bending mode is excited through a harmonic force applied perpendicularly at the tip of the blade. 
Its amplitude is $0.1 N$.

\subsection{Comparison between HBM, TCM and HDHB}
Eight nodes (green nodes on Fig.~\ref{fig:noeuds}) on each interface were selected.
Four different harmonic orders were calculated with the HBM, TCM and HDHB methods.
The number of time steps in the time domain was fixed at $Nit=105$ for trigonometric collocation and harmonic balance.
For the HDHB method this number was fixed by the order of harmonics $Nit=2*Nh+1$.
TCM and HBM were also tested with $Nit=2*Nh+1$ under the same conditions as the HDHB method.

\begin{figure}
\begin{center}
\includegraphics[width=0.9\columnwidth]{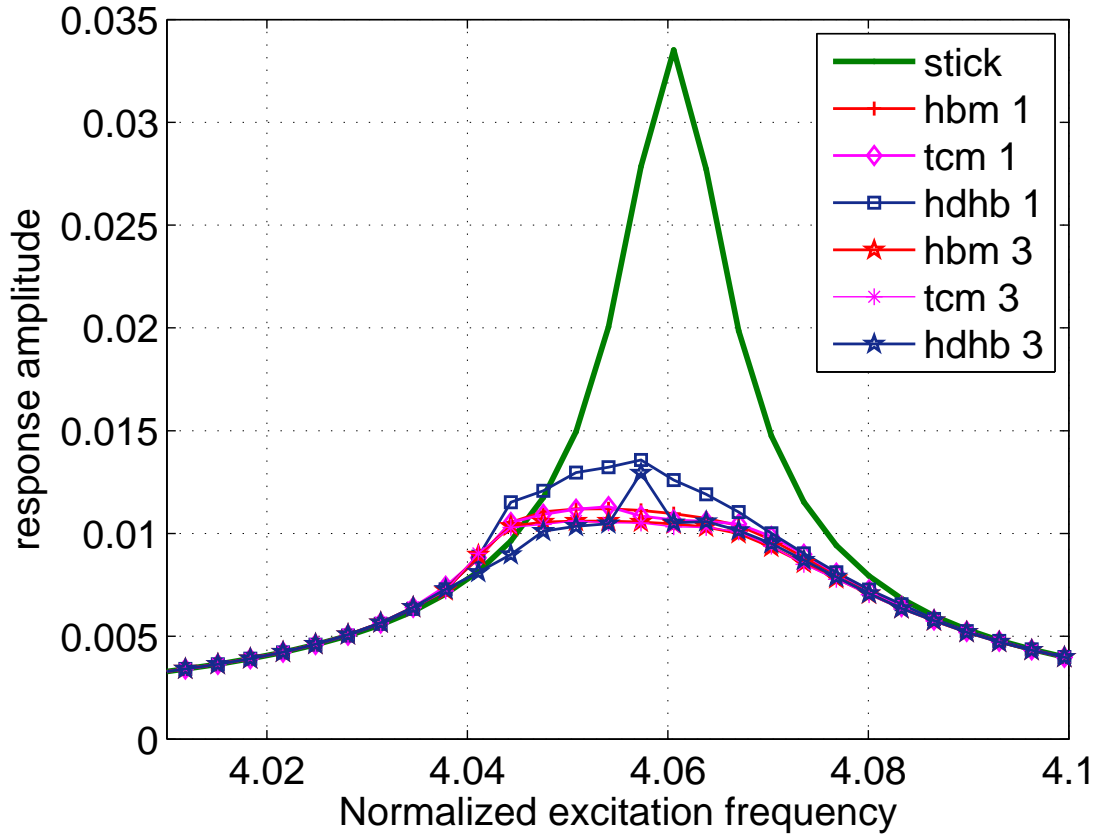}
      \label{fig:FRF_1_3}
\caption{Frequency response around first mode with different methods $Nh=1,3$}
\end{center}
\end{figure}

\begin{figure}
\begin{center}
   \includegraphics[width=0.9\columnwidth]{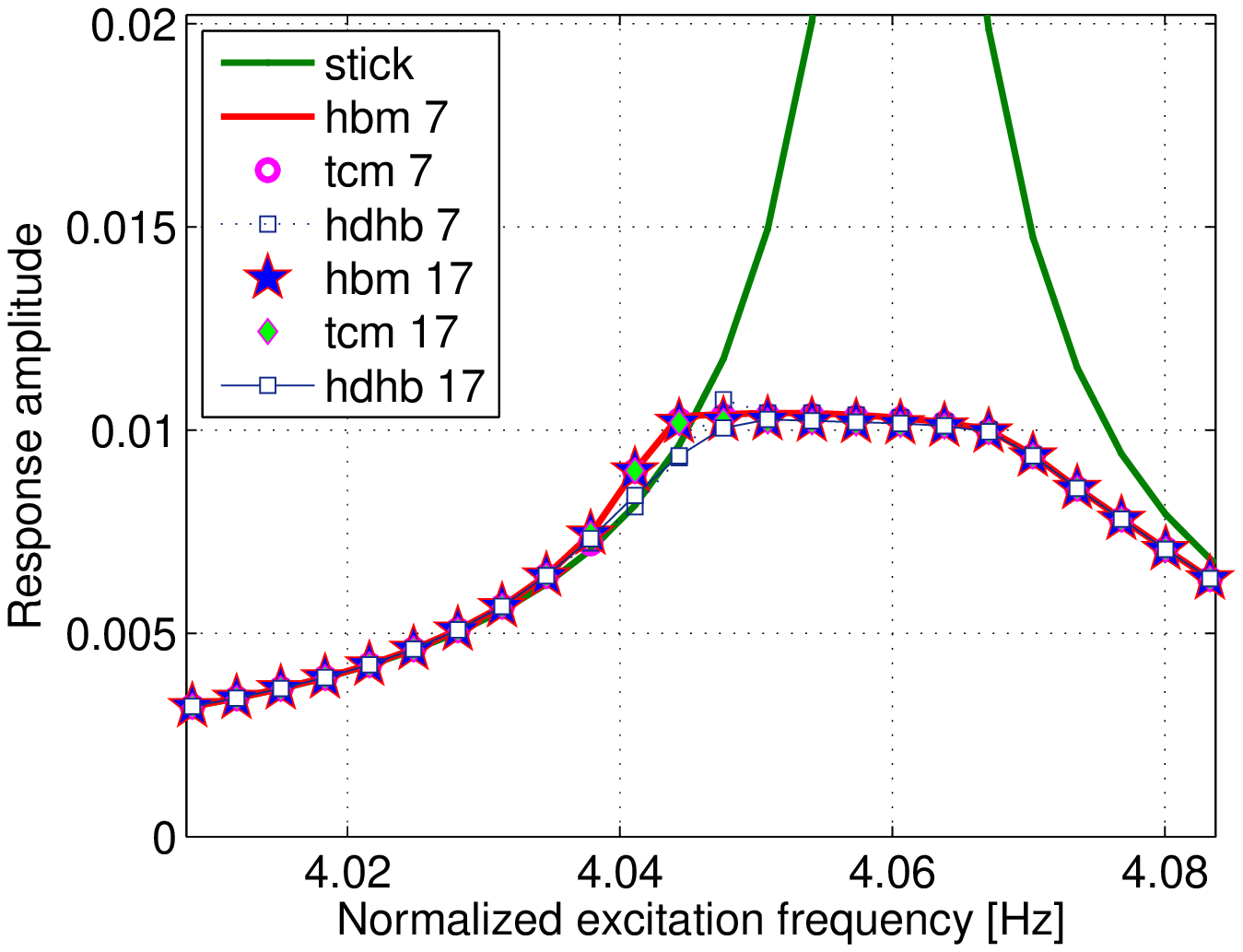}
      \label{fig:FRF_7_17}
\caption{Frequency response around first mode with different methods $Nh=7,17$}  
\end{center}
\end{figure}

\begin{figure}
    \begin{center}
        \includegraphics[width=0.9\columnwidth]{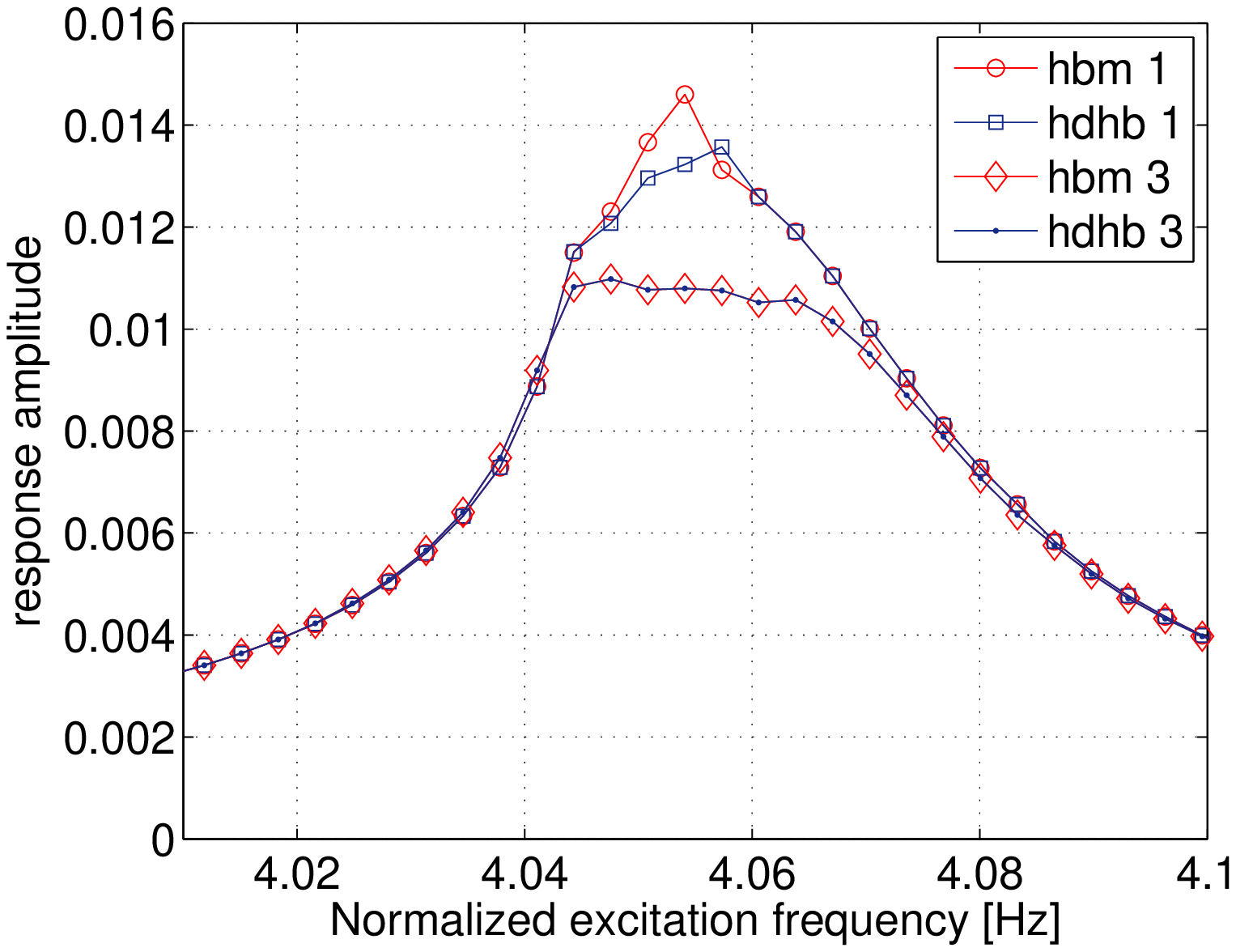}
    \end{center}
    \caption{Frequency response for HBM-1 with $Nit=2Nh+1$ and HDHB-1}
    \label{fig:comp_hb_1}
\end{figure}

The frequency responses are illustrated in Fig.~\ref{fig:FRF_1_3} and  Fig.~\ref{fig:FRF_7_17} for $Nh=1,3$ and $Nh=7,17$ respectively.
Fig.~\ref{fig:FRF_1_3} shows that HDHB gave results different from those of the HBM method for the low harmonic order.
This was caused by the low number of time steps in the time domain, therefore contact forces were poorly described, causing aliasing to occur.
As shown in Fig.~\ref{fig:comp_hb_1}, the same problem occured when $Nit=2*Nh+1$ in HBM.
When $Nh$ increased, the HDHB method gave better results and converged with HBM ( $Nit=105$).
This aspect of HDHB has already been observed by Liu~\cite{liu2006} for Duffing's oscillator.

\begin{figure}
    \begin{center}
        \includegraphics[width=0.9\columnwidth]{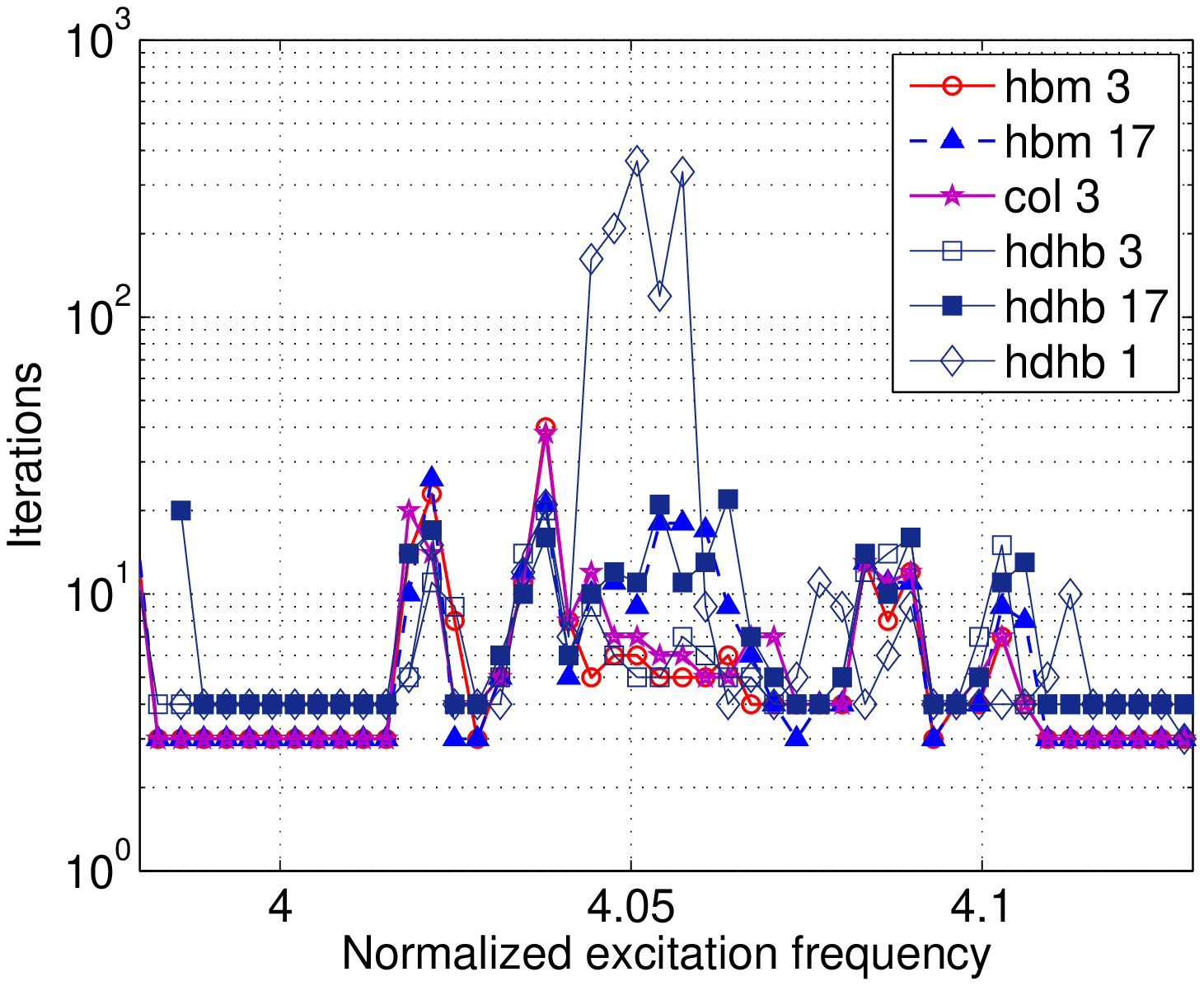}
    \end{center}
    \caption{Iterations versus frequency for different HB methods}
    \label{fig:it_freq}
\end{figure}

Table~\ref{tab:time} gives the CPU time for calculating $50$ frequencies of FRF with the different methods and for two time step numbers in the time domain. 
CPU time was normalized by CPU time for $Nh=1$ and $nit=105$.
Hybrid Powell nonlinear solver with trust-region dogleg method was used to solve non-linear algeraic system.
The number of iterations versus frequency is given in Fig.~\ref{fig:it_freq} to illustrate the difficulties of getting HDHB to converge for low harmonic order.
TCM requires slighty more iterations than HBM.
\begin{table*}\centering
\begin{tabular}{@{}lrrcrrcr@{}} \toprule
& \multicolumn{2}{c}{HBM} & \phantom{ab} & \multicolumn{2}{c}{TCM} & \phantom{ab} & \multicolumn{1}{c}{HDHB} \\ \cmidrule(r){2-3} \cmidrule(r){5-6} \cmidrule(r){8-8}
Nh & $nit=105$ & $nit=2Nh+1$ && $nit=105$ & $nit=2Nh+1$  && $nit=2Nh+1$  \\ \midrule
1  & 1       & 0.17          &&  1.43     & 0.18         && 0.17         \\
3  & 3.71    & 0.51          &&  2.70      & 0.45         && 0.71         \\
7  & 14.59   & 6.41          &&  7.10     & 3.24         && 3.71         \\
17 & 521     & 88.44         &&  44.62    & 46.60            && 46.70         \\
\bottomrule
\end{tabular}
\caption{CPU time necessary for calculating $50$ frequencies}
\label{tab:time}
\end{table*}

Good description of behaviour in the contact zone can be important for certain studies (fretting-wear for example).
Fig.~\ref{fig:comp_dep} shows the evolution of the tangential displacement of node 33 during one cycle at the resonance frequency for HBM-7, HDHB-7 and HDHB-17.
\begin{figure}
    \begin{center}
        \includegraphics[width=0.9\columnwidth]{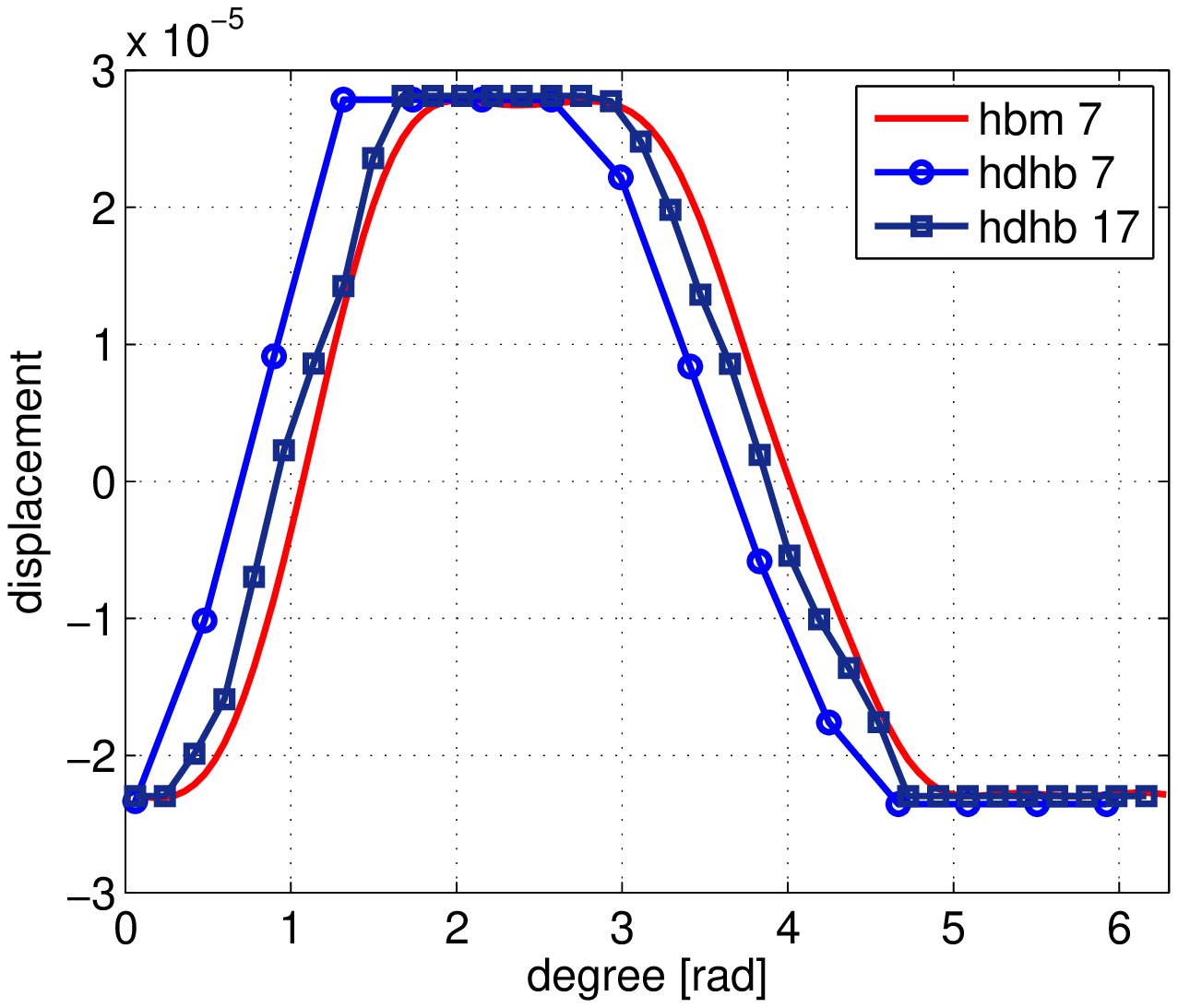}
    \end{center}
    \caption{Tangential displacement during one cycle}
    \label{fig:comp_dep}
\end{figure}
HBM and HDHB give fairly similar results and the amplitude is the same, but there is a phase difference between the two methods.
HDHB moves closer to the HBM curve at a higher harmonic order.

This study demonstrates that the three methods can be used to calculate FRF.
The HDHB method is better adapted to large systems with a high harmonic order as it is faster.

In the second part, we consider an example of fretting-wear in order to validate the pseudo-time method.

\subsection{Fretting-wear using the pseudo-time method}

\begin{figure}
    \begin{center}
        \includegraphics[width=0.9\columnwidth]{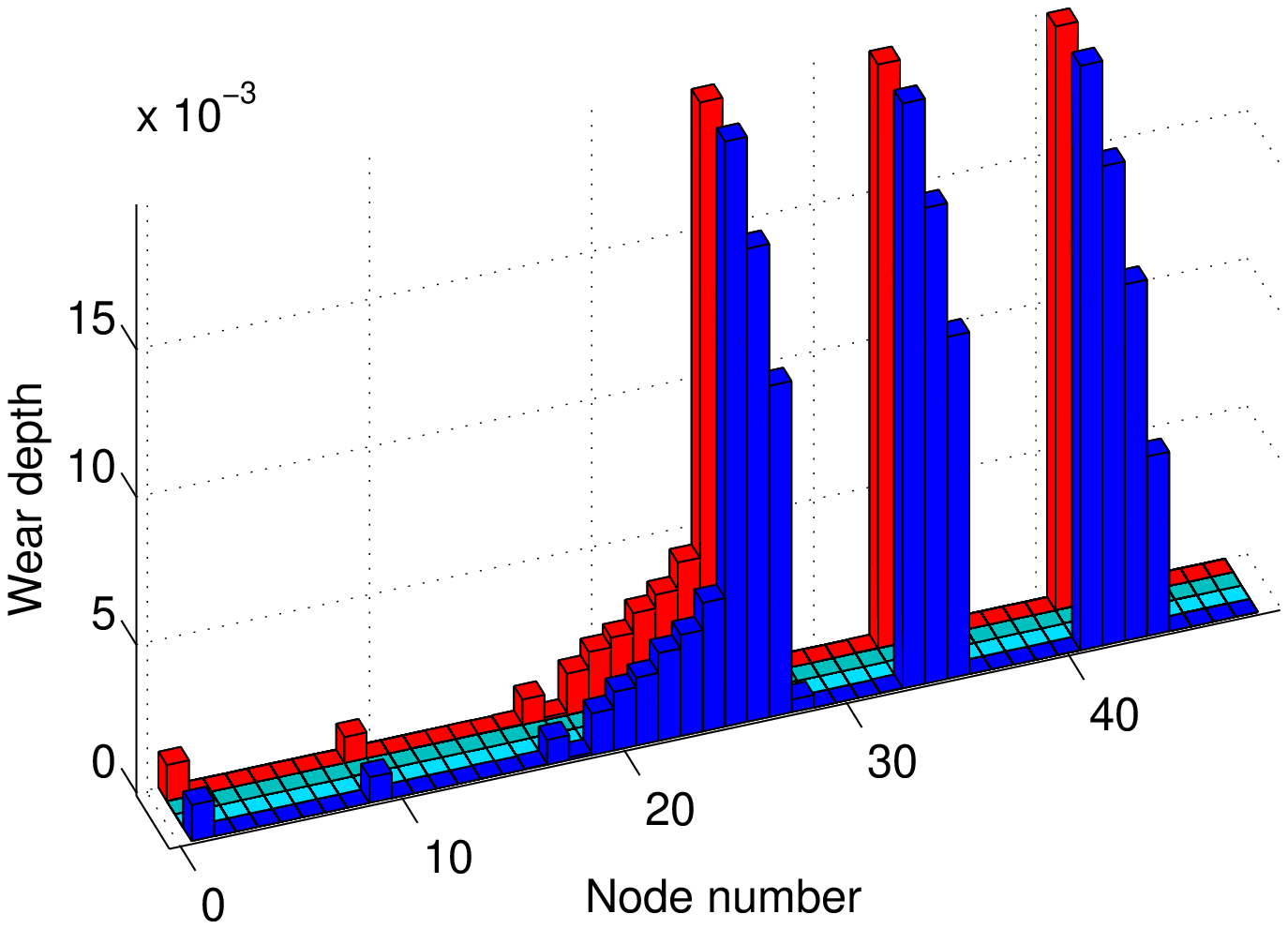}
            \end{center}
    \caption{Wear depth for each node in contact with the implicit scheme (foreground) and the pseudo-time method (background)}
    \label{fig:comp_Wf}
\end{figure}

\begin{figure}
    \begin{center}
        \includegraphics[width=0.9\columnwidth]{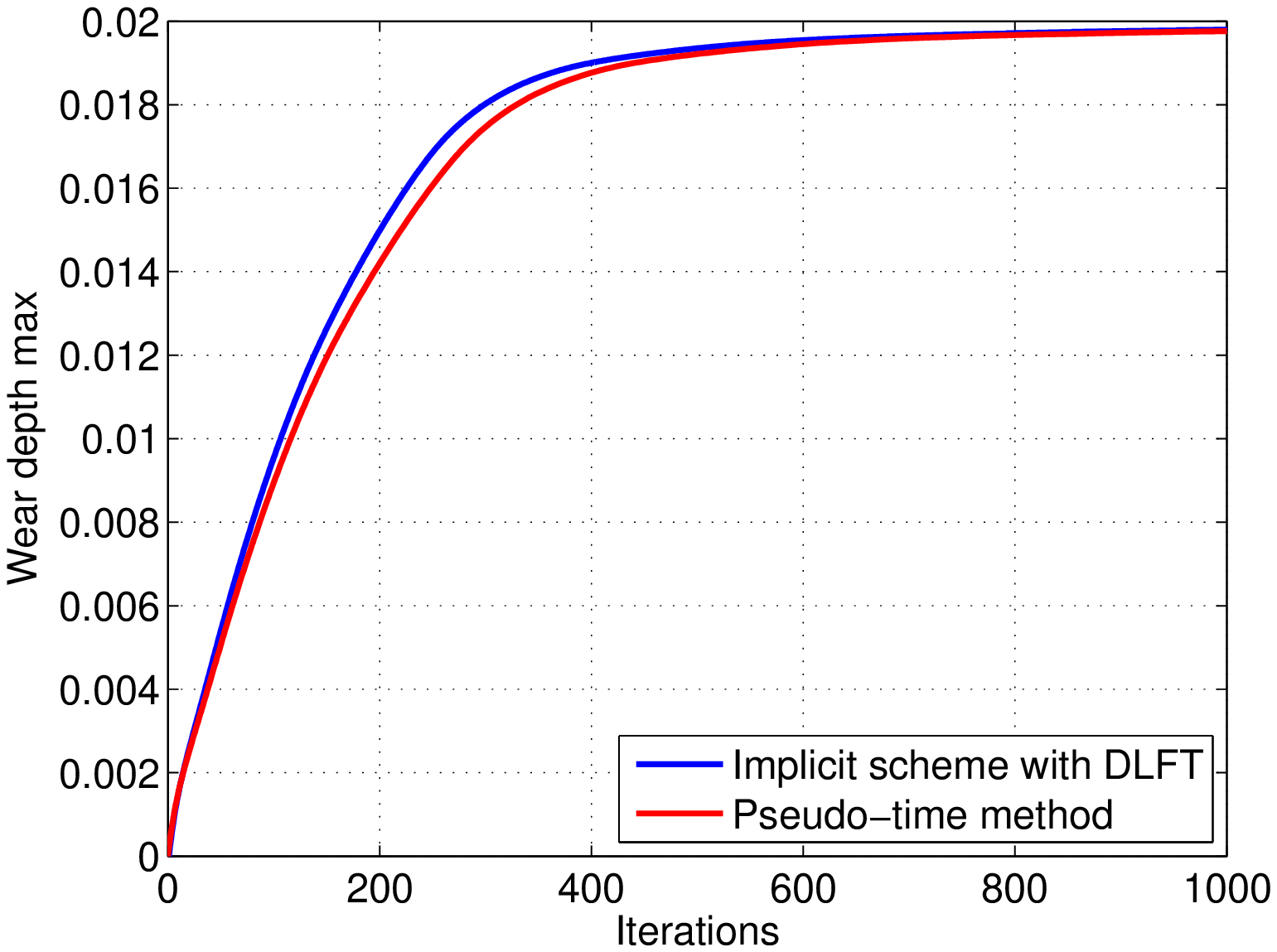}
    \end{center}
    \caption{Maximum wear depth evolution}
    \label{fig:it_W}
\end{figure}

\begin{figure}
    \begin{center}
        \includegraphics[width=0.9\columnwidth]{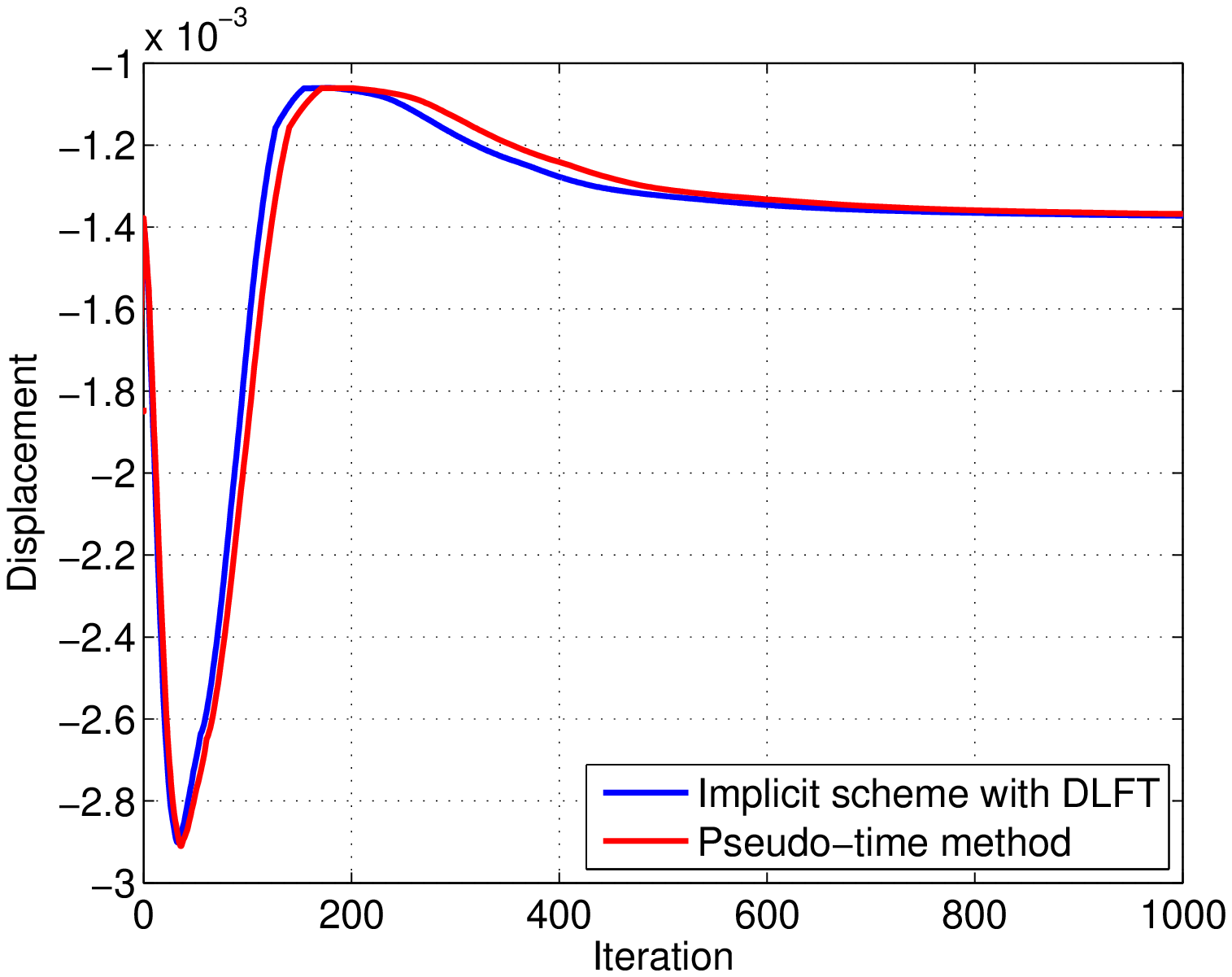}
    \end{center}
    \caption{Evolution of first harmonic value of tangential displacement for node 25}
    \label{fig:it_dep}
\end{figure}

The example is the same as that studied in a previous paper~\cite{salles2009}.
48 nodes on the contact surfaces and 3 harmonics were selected.
The amplitude of excitation was $Fe=0.5N$.
The wear profile was calculated by using an implicit scheme with the DLFT method for the non-linear dynamic problem, as proposed in~\cite{salles2009}, and also with the pseudo-time method that integrates wear kinetics in pseudo-time.
The results are shown in Fig.~\ref{fig:comp_Wf}.
The profiles are the same, thus the pseudo-time method works very well.
It is also interesting to know whether the dynamic behaviour at the tip of the blade is predicted well.
Fig~\ref{fig:comp_dep} shows the evolution of displacement at the tip of the blade during one cycle obtained by both methods. 
\begin{figure}
    \begin{center}
        \includegraphics[width=0.9\columnwidth]{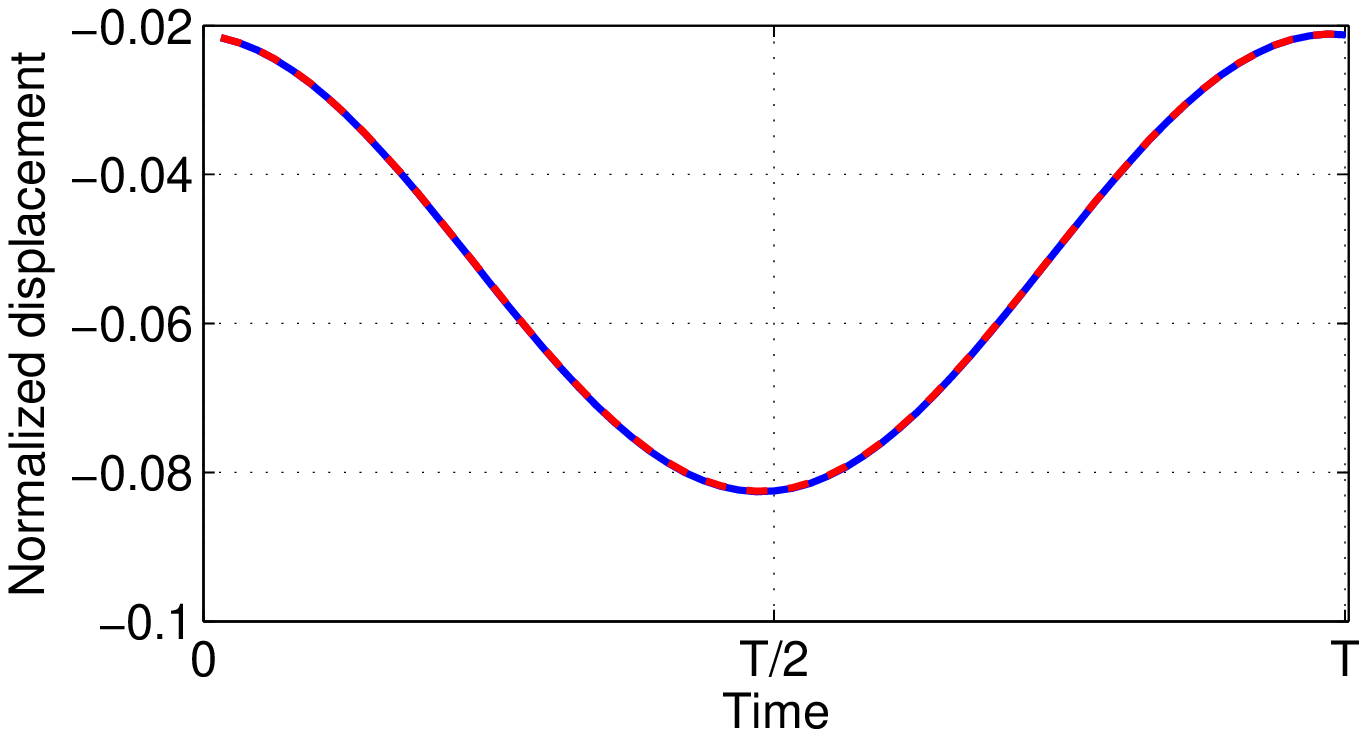}
    \end{center}
    \caption{Displacement at the tip of the blade}
    \label{fig:comp_dep}
\end{figure}
Moreover, the pseudo-time method takes less time to obtain the steady state.
The implicit method with DLFT requires 100\% CPU time while the pseudo time method requires 29\%.
Fig.~\ref{fig:it_W} and Fig.~\ref{fig:it_dep} illustrate the evolution of wear during the iterations.
In the implicit scheme, the iterations are directly linked with time.
This time in the pseudo-time method is non-physical. The same time scale is used for comparison. 
These figures highlight identical behaviour between methods regarding the evolution of wear and displacements.
The big advantage of the pseudo-time method is that the final steady state is obtained more rapidly than with the implicit scheme with DLFT.
On the other hand, the pseudo method cannot be applied to calculate transient wear evolution, since time is not physical.
We implemented this method in a MATLAB environment and tested different ode (ordinary differential equation) solvers.
The problem of fretting-wear under dynamical loadings is a difficult problem and thus explicit schemes converge badly.
Two implicit schemes work well: scheme based on Numerical Differentiation Formulae (NDF) and modified Rosenbrok Triple scheme~\cite{ode_suite1997} .
NDF scheme converges faster than other scheme and so it is the recommended scheme for pseudo-time method.

\section*{CONCLUSIONS AND PROSPECTS}

Three HB methods were introduced in this paper to perform simultaneous calculations of  non-linear vibrations of bladed disks.
Non-linearity is assumed at the contact with friction at the interface between disk and blade.
It was found that the HDHB method is well suited for large systems.
In the future, new contact algorithms will be developed on the basis that unknowns are variables in time, thereby making it possible to apply any static contact algorithm.

A new method was then introduced to solve the non-linear algebraic system by transforming it into ordinary differential equations whose solution is the steady state of the system.
Taking into account this approach we showed that the fretting-wear problem can be reformulated as a pseudo-time problem.
This approach is valid only if a steady state exists.
Numerical examples were used to show that both methods, i.e. the implicit scheme with a Newton solver and the pseudo-time method, give exactly the same results. 
Future works will be performed to optimize an ode solver based on the method proposed by Sicot~\cite{sicot2008}.

\bibliographystyle{asmems4}
\begin{acknowledgment}
We are grateful to Snecma for its technical and financial support. 
This work was carried out in the framework of the MAIA mechanical research and technology program sponsored by CNRS, ONERA and the SAFRAN Group.
\end{acknowledgment}

\bibliography{biblio_te2011}
\end{document}